\documentclass[useAMS]{mn2e}
\usepackage{graphicx}

\def\simgt{\ {\raise-.5ex\hbox{$\buildrel>\over\sim$}}\ }
\def\I{\'\i}
\def\cd{d$^{-1}$\,}
\def\kms{kms$^{-1}$\,}

\begin{document}

\title[Multisite photometry of V2052 Oph and V986 Oph]
{A multisite photometric study of two unusual $\beta$ Cep stars: 
the magnetic V2052 Oph and the massive rapid rotator V986 Oph}
\author[G. Handler et al.]
{G. Handler,$^{1}$ R. R. Shobbrook,$^{2}$ K. Uytterhoeven,$^{3,\,4}$ 
M. Briquet,$^{5,\,6}$ C. Neiner,$^{7}$ \and T. Tshenye,$^{8}$ B. 
Ngwato,$^{8}$ 
H. van Winckel,$^{6}$ E. Guggenberger,$^{9}$ G. Raskin,$^{6}$\and
E. Rodr\I guez,$^{10}$ A. Mazumdar,$^{11}$ C. Barban,$^{7}$ 
D. Lorenz,$^{9}$ B. Vandenbussche,$^{6}$ \and T. \c Sahin,$^{12,\,13}$ 
R. Medupe,$^{8}$ C. Aerts\,$^{6}$
\and \\
$^1$ Copernicus Astronomical Center, Bartycka 18, 00-716 Warsaw, Poland (gerald@camk.edu.pl)\\
$^{2}$ Research School of Astronomy and Astrophysics, Australian National University, Canberra, ACT, Australia\\
$^{3}$ Instituto de Astrof\'isica de Canarias (IAC), E-38200 La Laguna, 
Tenerife, Spain\\
$^{4}$ Departamento de Astrof\'isica, Universidad de La Laguna (ULL), 
E-38206 La Laguna, Tenerife, Spain\\
$^{5}$ Institut d'Astrophysique et de G\'eophysique, Universit\'e de 
Li\`ege, All\'ee du 6 Ao\^ut 17, B\^at B5c, 4000, Li\`ege, Belgium\\
$^{6}$ Instituut voor Sterrenkunde, K. U. Leuven, Celestijnenlaan 200D, 
B-3001 Leuven, Belgium\\
$^{7}$ LESIA, Observatoire de Paris, CNRS UMR 8109, UPMC, Universit\'e 
Paris Diderot; 5 place Jules Janssen, 92190 Meudon, France\\
$^{8}$ Department of Physics, University of the North-West, Private Bag
X2046, Mmabatho 2735, South Africa\\
$^{9}$ Institut f\"ur Astronomie, Universit\"at Wien, T\"urkenschanzstrasse
17, A-1180 Wien, Austria\\
$^{10}$ Instituto de Astrofisica de Andalucia, C.S.I.C., Apdo. 3004, 
18080 Granada, Spain\\
$^{11}$ Homi Bhabha Centre for Science Education (TIFR), V.\ N.\ Purav 
Marg, Mumbai 400088, India\\
$^{12}$ Akdeniz University, Faculty of Science, Space Science and
Technologies Department, 07058, Antalya, Turkey\\
$^{13}$ TUBITAK National Observatory, Akdeniz University Campus, 07058,
Antalya, Turkey}

\date{Accepted 2005 July 17.
 Received 2005 August 13;
 in original form 2005 September 10}
\maketitle

\begin{abstract}
We report a multisite photometric campaign for the $\beta$ Cep stars 
V2052 Oph and V986 Oph. 670 hours of high-quality differential 
photoelectric Str\"omgren, Johnson and Geneva time-series photometry 
were obtained with eight telescopes on five continents during 182 
nights. Frequency analyses of the V2052 Oph data enabled the detection 
of three pulsation frequencies, the first harmonic of the strongest 
signal, and the rotation frequency with its first harmonic. Pulsational 
mode identification from analysing the colour amplitude ratios confirms 
the dominant mode as being radial, whereas the other two oscillations 
are most likely $l=4$. Combining seismic constraints on the inclination 
of the rotation axis with published magnetic field analyses we conclude 
that the radial mode must be the fundamental. The rotational light 
modulation is in phase with published spectroscopic variability, and 
consistent with an oblique rotator for which both magnetic poles pass 
through the line of sight. The inclination of the rotation axis is 
$54\degr <i< 58\degr$ and the magnetic obliquity $58\degr <\beta< 
66\degr$. The possibility that V2052 Oph has a magnetically confined 
wind is discussed. The photometric amplitudes of the single oscillation 
of V986 Oph are most consistent with an $l=3$ mode, but this 
identification is uncertain. Additional intrinsic, apparently temporally 
incoherent, light variations of V986 Oph are reported. Different 
interpretations thereof cannot be distinguished at this point, but this 
kind of variability appears to be present in many OB stars. The 
prospects of obtaining asteroseismic information for more rapidly 
rotating $\beta$~Cep stars, which appear to prefer modes of higher $l$, 
are briefly discussed.
\end{abstract}


\section{Introduction}

For over a century, the $\beta$~Cep stars are known to be variable on 
time scales of hours (Frost 1902), but it took half a century longer to 
understand the nature of their variability, radial and nonradial 
pulsations (Ledoux 1951). Nowadays, about 300 members of this class of 
pulsating star are known (Stankov \& Handler 2005, Pigulski \& 
Pojma{\'n}ski 2008).

Because of the simultaneous presence of radial and nonradial oscillation 
modes in these stars, and their rather simple overall structure 
(basically a convective core and a radiative envelope), their potential 
as asteroseismic targets is evident. Asteroseismology is the inference 
of the interior structure of pulsating stars. This is accomplished by 
measuring their oscillation frequencies, comparing them with the 
eigenfrequencies of corresponding stellar models, and then fine-tuning 
those models to match the observed frequencies (see, e.g., Aerts et al.\ 
2010, Handler 2012).

Besides the Sun, the $\beta$~Cep stars were the first main sequence 
pulsators for which clear constraints on their inner structure could be 
obtained asteroseismically (for a heavily abbreviated literature, see 
Aerts et al.\ 2003, Pamyatnykh et al. 2004, Handler et al.\ 2009, Aerts 
et al.\ 2011). Results indicate that further increases in 
heavy-element opacities are needed, and some stars have been shown to 
rotate faster in their interior than on the outside.

These first successful studies were in most cases intentionally biased 
towards bright, slowly rotating stars. Slowly rotating $\beta$~Cep stars 
driven by the $\kappa$ mechanism tend to have higher pulsation 
amplitudes (Stankov \& Handler 2005) and therefore offer better 
possibilities for mode identification. Obviously, effects of rotation on 
the observed frequencies of axisymmetric modes of oscillation are also 
smaller, and rotationally split m-mode patterns would not overlap in 
frequency. This way of approaching asteroseismology of $\beta$~Cep 
stars proved to be sound. Therefore it appears reasonable to investigate 
targets that pose more difficult initial conditions, but that may also 
be more rewarding astrophysically.

V2052 Oph (HR 6684, $V=5.8$, B2IV-V) was discovered as a $\beta$ Cep 
pulsator by Jerzykiewicz (1972), and its dominant mode identified as 
radial (Heynderickx, Waelkens \& Smeyers 1994, Cugier, Dziembowski \& 
Pamyatnykh 1994). Neiner et al.\ (2003) carried out an extensive 
multiwavelength spectroscopic and spectropolarimetric study of V2052 Oph 
that revealed several interesting properties of this star. Besides the 
detection of a second, nonradial, pulsation mode, these authors could 
derive an accurate rotation period of $3.638833\pm0.000003$~d. V2052 Oph 
also possesses a dipole magnetic field. Based on new data of superior 
quality, Neiner et al.\ (2012a) determined $B_{\rm pol} \approx 400 G$, 
that the magnetic field is likely off-centred, and that He patches are 
present close to the magnetic poles. Because of the presence of a radial 
pulsation mode (that allows the determination of the mean stellar 
density), and of the known rotation period that makes it spin about 
twice as fast as the most "rapidly" rotating seismically well studied 
$\beta$~Cep star (12 Lac, Desmet et al.\ 2009), it was deemed worthwhile 
to devote a large observational effort to V2052 Oph. To this end, the 
present paper reports photometric results of a multisite campaign, 
whereas a companion paper (Briquet et al.\ 2012) deals with 
contemporaneous spectroscopy.

Located only a few degrees in the sky from V2052 Oph is another 
$\beta$~Cep star, V986 Oph (HR 6747, $V=6.1$, B0IIIn), with an 
interesting history in the literature. It is among the longest-period 
variables ($P \approx 0.29$~d, e.g., Jerzykiewicz 1975) of its class, 
and among the most luminous and hence most massive (Jones \& Shobbrook 
1974, Stankov \& Handler 2005). It is also a rapid rotator ($v \sin i = 
300$\,\kms, Abt, Levato \& Grosso 2002) and has been classified as a 
single-lined spectroscopic binary ($P_{\rm orb}=25.56$~d, $e=0.23$, 
Fullerton, Bolton \& Penrod 1985). Frequency analyses published by 
different authors indicate variability with periods between 7 to 8 
hours, but all studies noted that further photometric variability is 
present. However, no good explanation of its physical cause could be 
obtained (see Cuypers, Balona \& Marang (1989) for a detailed 
discussion). Furthermore, spectroscopic studies (Fullerton et al.\ 1985, 
Stateva, Niemczura \& Iliev 2010) implied that the short period 
variation is due to a mode of rather high spherical degree ($l=4$, 6 or 
8). V986 Oph was also photometrically monitored during this multisite 
campaign, in the hope to gain understanding of its variability.

\section{Observations and data reduction}

Our photometric observations were carried out at seven different 
observatories on five continents, from 11 March - 6 September 2004. An 
overview of the campaign observations is given in Table~1. In most 
cases, single-channel differential photoelectric photometry was acquired 
but at Sierra Nevada Observatory a simultaneous $uvby$ photometer was 
used. At observatories where no Str\"omgren $vy$ filters were available 
we used Johnson $V$, with Str\"omgren $u$ as possible complement. 
Finally, as the photometer at the Mercator telescope has Geneva filters 
installed permanently, we used this filter system. In the three seasons 
preceding this campaign, 72.7 hours of Geneva photometry had been 
obtained with the Mercator telescope. These are included here as well.

\begin{table*}
\caption[]{Log of the photometric measurements of V986 Oph and V2052 Oph
during the multisite campaign in 2004.
Observatories are ordered according to geographical longitude.}
\begin{center}
\begin{tabular}{lccccl}
\hline
Observatory & Telescope & \multicolumn{2}{c}{Amount
of data} & Filter(s) & Observer(s)\\
& & Nights & hours & & \\
\hline
T\"ubitak National Observatory, Turkey & 0.5m & 2 & 6.1 & V & TS\\
South African Astronomical Observatory & 0.5m & 13 & 50.3 & uvy & EG, BN \\
South African Astronomical Observatory & 0.75m & 7 & 30.5 & uvy & GH\\
South African Astronomical Observatory & 0.5m & 9 & 35.0 & uV & TT \\
Piszk\'estet\H o Observatory, Hungary & 0.5m & 3 & 9.2 & V & DL\\
Sierra Nevada Observatory, Spain & 0.9m & 4 & 19.7 & uvby & ER\\
Roque de los Muchachos Observatory, Spain & 1.2m Mercator & 49 & 155.2 & 
Geneva & KU, MB, HW, GR, AM, CB, BV\\
Fairborn Observatory, USA & 0.75m APT & 55 & 198.7 & uvy & $--$\\
Siding Spring Observatory, Australia & 0.6m & 40 & 168.9 & uvy & RRS\\
\hline
Total & & 182 & 673.6 \\
\hline
\end{tabular}
\end{center}
\end{table*}

We chose two comparison stars: HR 6689 ($V=5.96$, A3V) was already used 
as a comparison for V2052 Oph by Jerzykiewicz (1972, 1993), but was in 
the second paper suspected to be variable. HR 6719 ($V=6.34$, B2IV) was 
used as a comparison star by Jerzykiewicz (1993), but also suspected of 
variability. Subsequent photometric studies of the star did not mention 
variability, but Telting et al.\ (2007) reported line profile variations 
pointing towards pulsation of high azimuthal order. Although this choice 
of comparison stars may not seem ideal, we did not find better suited 
candidates in this part of the sky. Fortunately, the comparison 
stars proved to be constant within the accuracy of our measurements, and 
did not affect our results in any way. The targets were therefore 
observed alternatingly with these comparison stars, but V986 Oph was 
measured only in every other cycle.

Data reduction was started by compensating for coincidence losses and 
subtracting sky background. Extinction corrections had to be made in two 
steps caused by the two comparison stars always being located at 
systematically different air mass (similar right ascension, but 
different declination and close to the celestial equator). This means 
that even small errors in the applied extinction coefficients cause 
variations in the nightly photometric zeropoints.

Consequently, we first determined the extinction coefficients with the 
standard Bouguer method from the comparison star measurements. We then 
examined the differential comparison star light curves for variability, 
resulting in a non-detection. Next, we imposed that the average nightly 
photometric zeropoints for each instrumental system be the same and 
correspondingly amended the extinction corrections within reasonable 
limits. This procedure considerably improved the accuracy of our final 
light curves, as examined with the target star data. The residual 
scatter in the differential comparison star data is between 4.6 and 3.5 
mmag in the Str\"omgren filters, and between 3.1 and 2.4 mmag in the 
Geneva measurements.

Consequently, we computed differential light curves of the target stars 
and heliocentrically corrected their timings. The single-colour 
measurements were binned to sampling intervals similar to that of the 
multicolour measurements to avoid giving them higher weight in the 
consequent analyses. Finally, the photometric zeropoints of the 
different instruments were compared between the different sites and 
adjusted if necessary. The resulting final combined time series, 
spanning 179.3~d, was subjected to frequency analysis. Light curves from 
the central part of our campaign are shown in Fig.\ 1, together with 
fits to be derived and commented on in what follows.

\begin{figure*}
\includegraphics[angle=270,width=180mm,viewport=0 3 270 579]{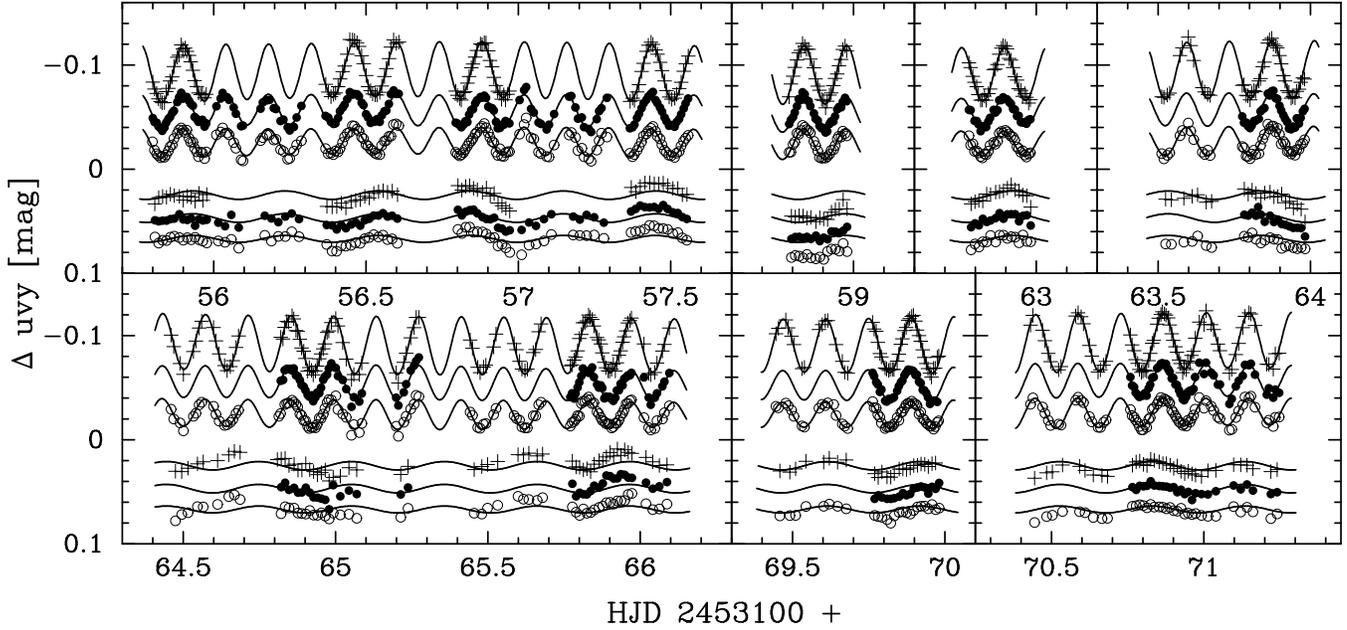}
\caption[]{Some of our time-series photometry of V2052 Oph (upper three 
curves) and V986 Oph (lower three curves). The plus signs represent $U$ 
or $u$ measurements, the filled circles are $v$ data and the open 
circles data in the $V$ or $y$ filter. The lines are the multifrequency 
fits derived from periodicity search.}
\end{figure*}

\section{Frequency analysis}

The heliocentrically corrected data were searched for periodicities 
using the program {\tt Period04} (Lenz \& Breger 2005). This package 
applies single-frequency power spectrum analysis and simultaneous 
multi-frequency sine-wave fitting. It also includes advanced options 
such as the calculation of optimal light-curve fits for multiperiodic 
signals including harmonic and combination frequencies.

For purposes of frequency detection, the Str\"omgren $u$ and Geneva $U$ 
filter data were merged after checking that the oscillation amplitudes 
were the same within the errors. Measurements in the Str\"omgren $y$ and 
Johnson and Geneva $V$ filters were treated as equivalent due to the 
same effective wavelength of these filters, and were analysed together. 
After signals were believed to be detected, their presence was checked 
in the data of the individual filters (that is, the seven Geneva 
filters, Str\"omgren $uv$, and the combined Str\"omgren $y$/Johnson $V$ 
light curves). The Str\"omgren $b$ filter measurements were not used 
because too few data are available.

Amplitude spectra were computed, compared with the spectral window
functions, and the frequencies of the intrinsic and statistically
significant peaks in the Fourier spectra were determined. Multifrequency
fits with all detected signals were calculated step by step, the
corresponding frequencies, amplitudes and phases were optimized and
subtracted from the data before computing residual amplitude spectra,
which were then examined in the same way.

We consider an independent peak statistically significant if it exceeds an
amplitude signal-to-noise ratio of 4 in the periodogram; combination
signals must satisfy $S/N>3.5$ to be regarded as significant (see Breger
et al.\ 1993, 1999). The noise level was calculated as the average
amplitude in a 5 \cd interval centred on the frequency of interest.

For the detection of pulsation frequencies we did not make use of the
pre-campaign Geneva measurements as those were carried out without using
comparison stars. They therefore have about a factor of 2.5 higher scatter
than the campaign data and increase the noise level in a joint analysis.
However, the pre-campaign measurements could in some cases be used to
derive more precise frequency values.

\subsection{V2052 Oph}

We started by computing the Fourier spectral window of the data, which
turned out reasonably clean. The strongest aliases in the u/U and y/V data
have only 36\% of the amplitude of the true signal. The amplitude spectrum 
of the data itself, dominated by the known radial mode frequency, is shown 
in the upper panel of Fig.\ 2. We chose to use the u/U data for 
presentation purposes for the pragmatical reason that all signals to be 
reported are detected in this data set alone.

\begin{figure}
\includegraphics[width=85mm,viewport=0 0 276 454]{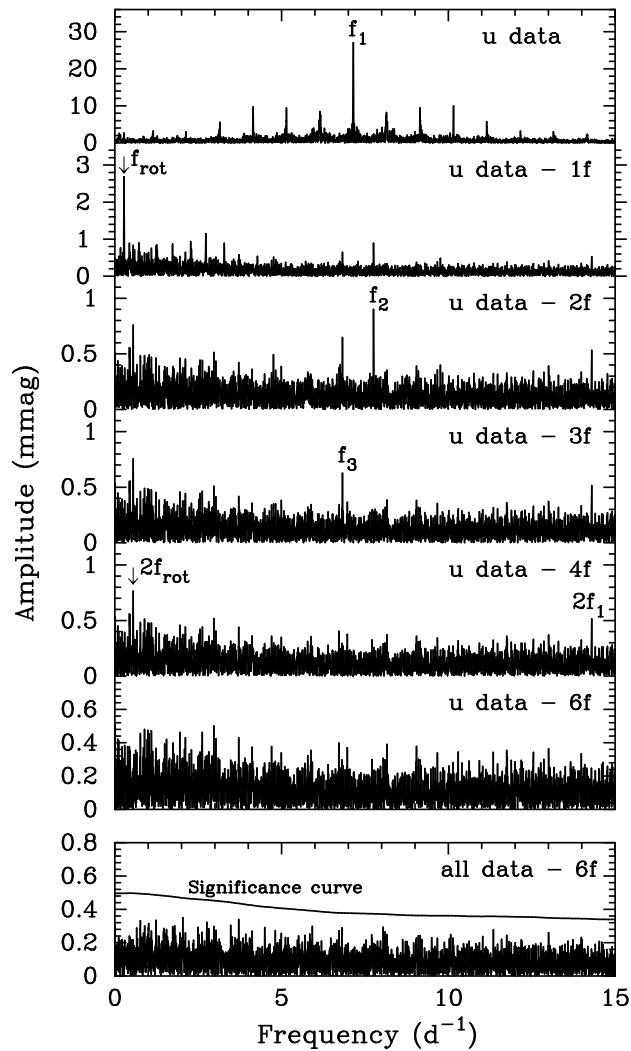}
\caption[]{Amplitude spectra of our combined $u/U$ filter data of V2052 
Oph with consecutive prewhitening steps. The uppermost panel is 
practically identical with the spectral window of the data. The lowest 
panel shows the amplitude spectrum of the combined residuals in all 
filters.} 
\end{figure}

Prewhitening the strongest signal from the data and examining the 
residual amplitude spectrum, we recover the stellar rotation frequency 
(second panel of Fig.\ 2). Further analysis reveals two more signals in 
the frequency domain of the radial mode, as well as the first harmonic 
of the first mode and of the rotation frequency. The residual amplitude 
spectrum after prewhitening these six frequencies shows a slight 1/f 
component, as expected from residual atmospheric effects in the data, 
and no signal in excess of 0.5 mmag.

With frequency solutions for the individual filters as starting values, 
we attempted to improve the accuracy of our frequency determinations by 
including the pre-campaign observations, therefore increasing the time 
base of the data set by a factor of 6.5. By examining the u/U and y/V 
data as well as confronting the results, we obtained more accurate 
values for all frequencies, with the exception of the second pulsation 
frequency where we encountered an aliasing problem. Tests on which value 
would result in lower residuals etc.\ did not allow to determine a 
preferred value, and the choice of this frequency did not affect the 
outcome on the others. We therefore adopted the average of the two 
candidate values, and used half the alias spacing as its uncertainty. 
The final values of the frequencies were then fitted to the campaign 
data alone but kept fixed, and only the amplitudes, phases and zeropoint 
were left as free parameters. The result of this procedure is listed in 
Table 2.

\begin{table*}
\caption[]{Multifrequency solution for our time-resolved photometry of
V2052 Oph. Error estimates for the independent frequencies were derived 
from considering both formal errors (following Montgomery \& O'Donoghue 
1999) and differences in the u/U and y/V results. The quoted 
errors on the amplitudes are the formal values. The S/N ratio is for the 
$u$ filter data.}
\begin{center}
\begin{tabular}{lcccccc}
\hline
ID & $f_{1}$ & $f_{2}$ & $f_{3}$ & 2$f_{1}$ & $f_{rot}$ & $2f_{rot}$ \\
\hline
Frequency (\cd) & 7.148474 $\pm$ 0.000005 &  7.7567 $\pm$ 0.0007 & 
6.82216 $\pm$ 0.00005 &  14.296948 & 0.27480 $\pm$ 0.00002   & 0.54960 \\
$u$ Ampl.\ (mmag) & 26.98 $\pm$ 0.10 & 0.84 $\pm$ 0.10 & 0.54 $\pm$ 0.10 
& 0.52 $\pm$ 0.10 & 2.76 $\pm$ 0.10 & 0.78 $\pm$ 0.10\\
$v$ Ampl.\ (mmag) & 15.41 $\pm$ 0.09 & 0.85 $\pm$ 0.09 & 0.60 $\pm$ 0.09 
& 0.20 $\pm$ 0.09 & 1.97 $\pm$ 0.09 & 1.05 $\pm$ 0.09 \\
$y$ Ampl.\ (mmag) & 13.37 $\pm$ 0.08 & 0.90 $\pm$ 0.08 & 0.56 $\pm$ 
0.08 & 0.23 $\pm$ 0.08 & 1.62 $\pm$ 0.08 & 0.29 $\pm$ 0.08 \\
$U$ Ampl.\ (mmag) & 26.65 $\pm$ 0.16 & 0.93 $\pm$ 0.16 & 0.97 $\pm$ 0.16 & 
0.53 $\pm$ 0.16 & 2.89 $\pm$ 0.16 & 0.47 $\pm$ 0.16 \\
$B_1$ Ampl.\ (mmag) & 16.08 $\pm$ 0.15 & 0.78 $\pm$ 0.15 & 0.97 $\pm$ 0.15 
& 0.26 $\pm$ 0.15 & 2.56 $\pm$ 0.15 & 1.35 $\pm$ 0.15\\
$B$ Ampl.\ (mmag) & 15.19 $\pm$ 0.15 & 0.81 $\pm$ 0.15 & 0.75 $\pm$ 0.15 
& 0.27 $\pm$ 0.15 & 2.21 $\pm$ 0.15 & 0.91 $\pm$ 0.15 \\
$B_2$ Ampl.\ (mmag) & 14.40 $\pm$ 0.16 & 0.67 $\pm$ 0.16 & 0.60 $\pm$ 
0.16 & 0.07 $\pm$ 0.16 & 1.88 $\pm$ 0.16 & 0.34 $\pm$ 0.16 \\
$V_1$ Ampl.\ (mmag) & 13.18 $\pm$ 0.15 & 0.90 $\pm$ 0.15 & 0.76 $\pm$ 
0.15 & 0.38 $\pm$ 0.15 & 1.47 $\pm$ 0.15 & 0.29 $\pm$ 0.15\\
$V$ Ampl.\ (mmag) & 13.22 $\pm$ 0.13 & 0.96 $\pm$ 0.13 & 0.74 $\pm$ 0.13 
& 0.29 $\pm$ 0.13 & 1.40 $\pm$ 0.13 & 0.32 $\pm$ 0.13 \\
$G$ Ampl.\ (mmag) & 12.83 $\pm$ 0.16 & 1.04 $\pm$ 0.16 & 0.60 $\pm$ 0.16 & 
0.41 $\pm$ 0.16 & 1.56 $\pm$ 0.16 & 0.27 $\pm$ 0.16 \\
$S/N$ & 200.2 & 6.4 & 4.0 & 3.9 & 14.1 & 4.0\\
\hline
\end{tabular}
\end{center}
\end{table*}

The multifrequency fit listed in the table represents the data within rms
residuals between 3.4 to 2.7 mmag (Str\"omgren data) and between 2.6 to
2.2 mmag (Geneva data). To search for possible additional signals, we
merged the residual data from all filters and computed the combined 
amplitude spectrum (lowest panel of Fig.\ 2). It contains no peak in 
excess of 0.35 mmag, and none with $S/N \geq 3.2$.

Examining the wavelength dependence of the phases of the independent
signals, we noticed that the dominant pulsation signal is not in phase in
all filters, as demonstrated in Fig.\ 3. In particular,
$\phi_v-\phi_y=0.9\pm0.3\degr$ and $\phi_u-\phi_y=4.2\pm0.2\degr$, i.e.\
the shorter the wavelength, the later light maximum/minimum is reached. As
the amplitudes of the other two signals in this frequency range are by at
least a factor of 15 smaller, the errors in the phases are correspondingly
larger. Consequently, no statistically significant phase shifts within the
different filter passbands have been detected for $f_2$ and $f_3$.

\begin{figure}
\includegraphics[width=85mm,viewport=0 00 269 167]{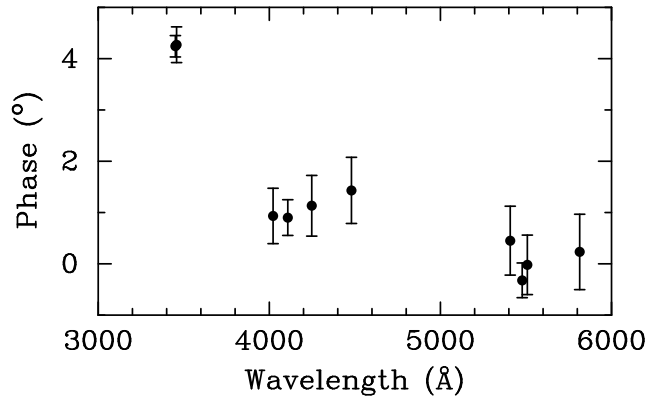}
\caption[]{Phase of the dominant oscillation mode of V2052 Oph with 
respect to that in the Str\"omgren y filter. The dots with the error bars 
are the measured phase shifts in all Geneva and Str\"omgren filters.} 
\end{figure}

The Fourier parameters of the rotational light variation change 
substantially from filter to filter (cf.\ Table 2). To determine its 
shape we first removed the pulsational variability from the data, and 
then phased them with respect to the rotation period. The Geneva data 
were summed into 20 phase bins and the more numerous Str\"omgren 
measurements in 25 bins. The rotational light curves are shown in Fig.\ 
4. Unfortunately, these cannot be compared with counterparts from 
satellite missions such as Kepler and CoRoT as our passbands are not 
sufficiently red sensitive.

\begin{figure}
\includegraphics[width=85mm,viewport=0 02 259 558]{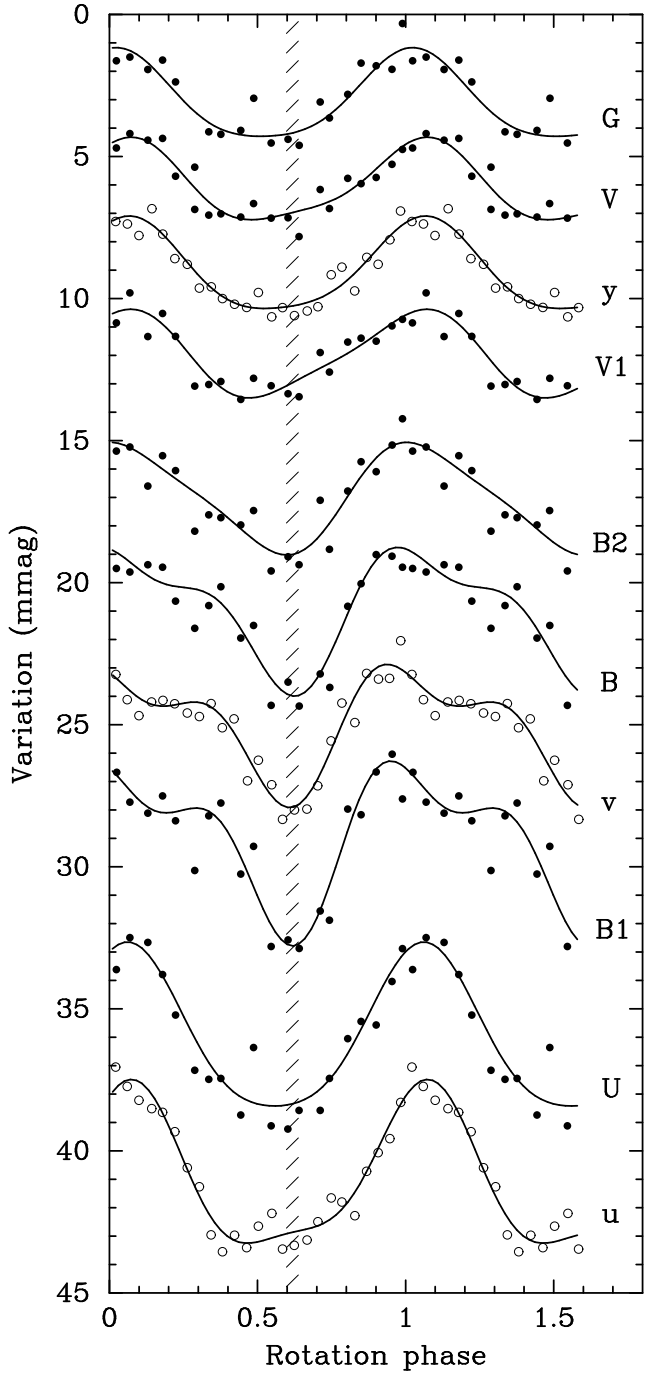}
\caption[]{Phase diagrams of the rotational light variations of 
V2052 Oph in the different filters, plotted according to decreasing 
effective wavelength from top to bottom. Str\"omgren data are shown as 
open circles, Geneva data as filled circles, and fits are overlaid to 
guide the eye. The hatched area denotes the rotational phase zero as 
defined by Neiner et al.\ (2003).}
\end{figure}

While we have arbitrarily phased these light curves and fits relative to
HJD 2453000.000, the hatched area in Fig.\ 4 indicates the phase of
minimum equivalent width of the ultraviolet spectral lines studied by
Neiner et al.\ (2003). We will return to discuss this phasing in Sect.\ 
5.1.

\subsection{V986 Oph}

The frequency analysis of our photometry of V986 Oph was performed in a 
similar way to that of V2052 Oph, and we also choose the u/U data for 
presentation. The amplitude spectrum of V986 Oph appears simple, with 
only one significant frequency present (Fig.\ 5). However, the residuals 
left behind a single-frequency solution are between 6.3 and 7.2 mmag per 
point in the Str\"omgren data, and between 7.0 and 8.1 mmag per point in 
the Geneva data, thus about a factor of two to three higher than those 
for V2052 Oph or for the differential comparison star data. The poorer 
fit for V986 Oph is also readily visible in Fig.\ 1. The noise level in 
the residual amplitude spectrum of V986 Oph is even about a factor of 
five higher because of the smaller amount of data points available. 
Furthermore, the single frequency is not significantly detected in the 
Geneva data alone, neither in those obtained during the campaign, nor in 
the pre-campaign observations.

\begin{figure}
\includegraphics[width=85mm,viewport=0 0 258 223]{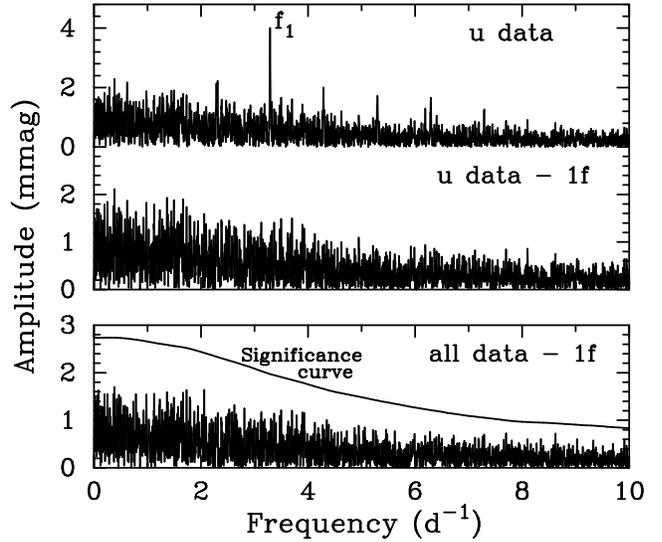}
\caption[]{Amplitude spectra and prewhitening of our combined $u/U$ filter 
data of V986 Oph.}
\end{figure}

Investigating the matter deeper and keeping in mind that spectroscopic 
binarity of V986 Oph has been reported, we first looked for a possible 
light time effect. We therefore merged the u/U, v and y/V data into bins 
no larger than 2.5~d as a compromise between not undersampling the 
reported 25.56~d orbit and having enough data points to determine the 
phase of the main light variation. We did not find a statistically 
significant light time effect, within a limit of 1270\,s at the orbital 
period. In one night of observation (around HJD 2453158.9, see Fig.\ 1) 
a $\sim0.02$~mag drop in light, suspicious of an eclipse, was present, 
but another data set obtained one prospective orbital period later did 
not show such a feature.

Looking at the pulsation amplitude now, it appears that it dropped 
somewhat during the course of the campaign, but not exceeding the 
$2\sigma$ level.  We therefore assumed a constant amplitude for the 
remainder of this work, calculated the frequency of the single 
significant signal as a S/N-weighted average in the different 
Str\"omgren filters, and determined its amplitude and phase in all 
filters (see Table 3). The signal was found to be in phase within the 
errors in all passbands.

\begin{table}
\caption[]{Frequency solution for our time-resolved Str\"omgren photometry 
of V986 Oph. The error on the frequency was determined from considering 
both formal errors (following Montgomery \& O'Donoghue 1999) and 
differences between the u, v, and y frequency solutions. The quoted errors 
on the amplitudes are the formal values. The S/N ratio is for the u 
filter data.}
\begin{center}
\begin{tabular}{lc}
\hline
ID & $f_{1}$ \\
\hline
Frequency (\cd) & 3.2886 $\pm$ 0.0003\\
$u$ Ampl.\ (mmag) & 4.0 $\pm$ 0.3\\
$v$ Ampl.\ (mmag) & 4.1 $\pm$ 0.3\\
$y$ Ampl.\ (mmag) & 3.3 $\pm$ 0.2\\
$S/N$ & 7.1\\
\hline
\end{tabular}
\end{center}
\end{table}

\section{Mode identification}

We now attempt to identify the spherical degree $l$ of the pulsation
modes by means of the $uvy$ and Geneva passband amplitudes of the
pulsational signals detected in the light curves. These amplitudes are to
be compared with theoretically predicted ones from model computations,
requiring the model parameter space to be constrained first. In other
words, we need to determine the positions of the two target stars in the
HR diagram as a starting point.

\subsection{V2052 Oph}

The latest spectroscopic $T_{\rm eff}$/log~$g$ values for V2052 Oph 
originate from Morel et al. (2006): $T_{\rm eff} = 23000 \pm 1000$~K and 
log~$g=4.0\pm0.2$, who also list $v \sin i = 61$\,\kms (including the 
macroturbulence velocity). Niemczura \& Daszy{\'n}ska-Daszkiewicz (2005) 
derived $T_{\rm eff} = 23300 \pm 700$~K and log~$g=3.89$ from 
low-resolution ultraviolet spectra. 

The online version of The General Catalogue of Photometric Data (GCPD; 
Mermilliod, Mermilliod \& Hauck 1997) contains standard Str\"omgren and 
Geneva photometric colours for the star. The Str\"omgren system 
calibration by Napiwotzki, Sch\"onberner \& Wenske (1993), yields 
$T_{\rm eff} = 22700 \pm 900$\,K, log~$g=3.9 \pm 0.3$, and also provides 
an absolute magnitude estimate using the calibration of Balona \& 
Shobbrook (1974): $M_v=-2.59$. The model atmosphere calibration of the 
Geneva system (K\"unzli et al.\ 1997) gives $T_{\rm eff} = 22800 \pm 
500$\,K, log~$g=3.8\pm0.3$. A relatively accurate HIPPARCOS parallax 
(van Leeuwen 2007) is also available: $\pi = 2.40 \pm 0.41$\,mas. 
Adopting $E(b-y)=0.230$ from Str\"omgren photometry, this leads to 
$M_v=-3.3^{+0.4}_{-0.3}$.

All these individual determinations are in very good agreement. We 
therefore assume $T_{\rm eff} = 22900 \pm 1000$~K and log~$g=4.0\pm0.2$. 
The tables by Flower (1997) then provide $BC=-2.20$. A comparison of the 
$T_{\rm eff}$/log~$g$ values with model evolutionary tracks prefers the 
lower value of our two absolute magnitude estimates, and 
suggests $M=9.2^{+0.9}_{-0.7}$~M$_{\odot}$.

Therefore we computed theoretical photometric amplitudes of the $0 
\leq l \leq 7$ modes for models with masses between 8.5 and 10.0 
$M_{\sun}$ in steps of 0.5 $M_{\sun}$, in a temperature range of 
$4.340 \leq \log T_{\rm eff} \leq 4.379$. We used OP opacities (e.g., 
Seaton 2005) and the Asplund et al.\ (2004) mixture. An overall metal 
abundance $Z=0.012$ and a hydrogen abundance of $X=0.7$ has been 
adopted, and no convective core overshooting was used. We are aware that 
Morel et al.\ (2006) and Niemczura \& Daszy{\'n}ska-Daszkiewicz (2005) 
derived a somewhat lower metallicity, but this small inconsistency is 
not crucial in the mode identification process.

We extracted theoretically calculated nonadiabatic parameters from
the models to determine the amplitudes in the different wavebands. This
approach follows that by Balona \& Evers (1999) and uses the same
software, hence we refer to this paper for details on the procedure.
Consequently, we computed the ratios of the theoretical amplitudes with
respect to those in the Str\"omgren u filter, for modes of spherical
degree $0 \leq l \leq 7$ and frequencies between 6.3 and 8.3 \cd, and
compared them with the observations (left-hand side panels of Fig.\ 6).

\begin{figure*}
\includegraphics[width=180mm,viewport=-33 00 533 462]{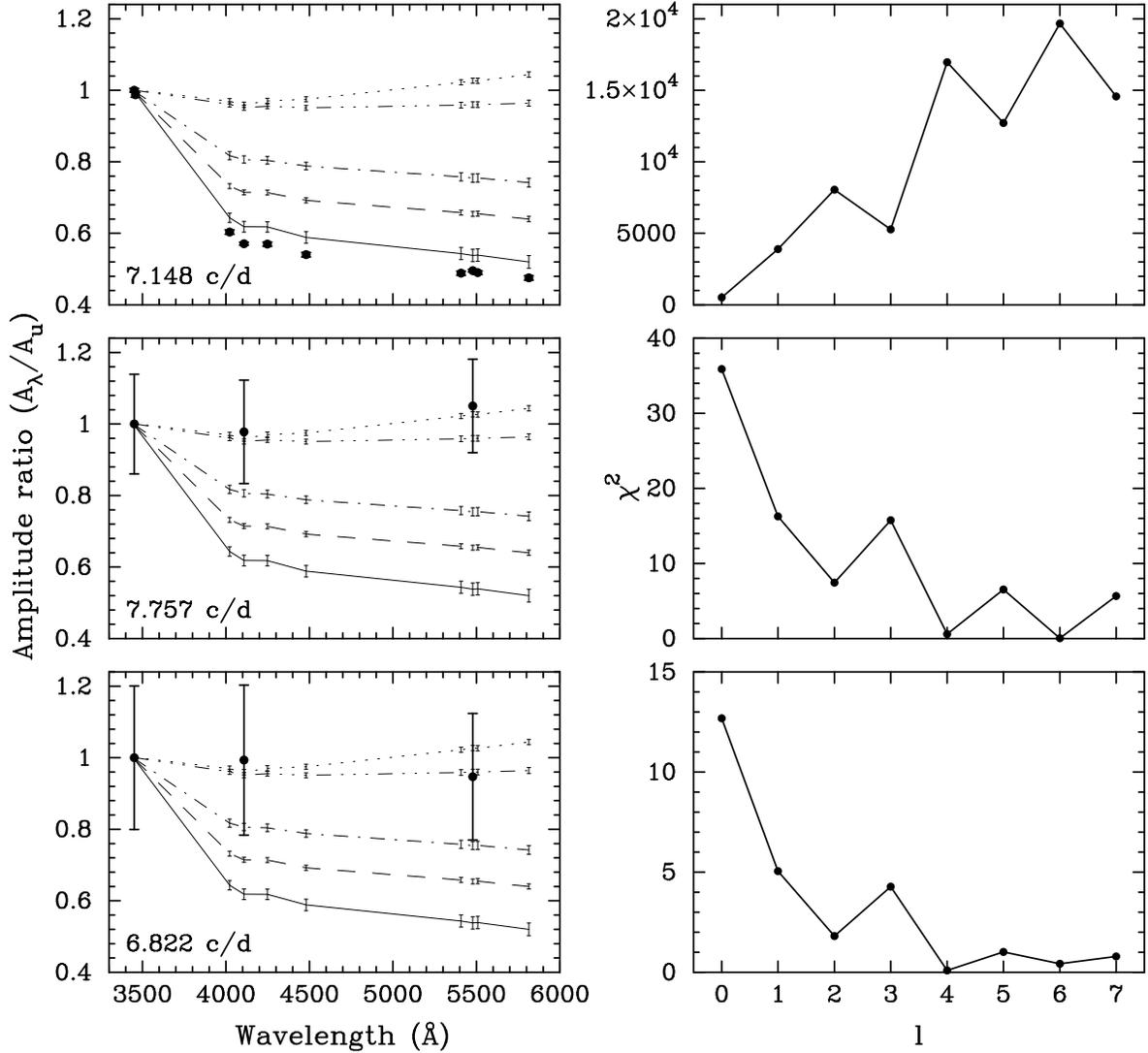}
\caption[]{Mode identifications for V2052 Oph from a comparison of
observed and theoretical amplitudes in the Str\"omgren and Geneva bands.
Left panels: amplitude ratios, normalised to unity at $u$. The filled
circles with error bars are the observed amplitude ratios, the thin error
bars denote the uncertainties in the theoretical amplitude ratios. The
full lines are theoretical predictions for radial modes, the dashed lines
for dipole modes, the dashed-dotted lines for quadrupole modes, the
dashed-triple-dotted lines are for $l=4$, and the dotted lines for
$l=6$ modes. The theoretical amplitude ratios for $l=3, 5$ and 7 are
not shown to avoid overcrowding. Right panels: $\chi^2$ analysis of the
photometric amplitudes. The smaller $\chi^2$, the more likely an
identification is.}
\end{figure*}

The right-hand side of Fig.\ 6 shows a $\chi^2$ analysis of the
photometric amplitudes, as outlined by Handler, Shobbrook \& Mokgwetsi
(2005). These $\chi^2$ values use the measurements and standard
errors of the amplitudes normalized to the mean of all passbands only, as
the pulsation phases carry no additional information on the mode type in 
our case.

Because of the high $S/N$ of the dominant mode of V2052 Oph, the 
amplitude ratios in all the individual Str\"omgren and Geneva filters 
could be incorporated. Such an approach is not optimal for the two 
low-amplitude pulsation modes. For their identification, we considered 
the combined $u/U$, $y/V$ as well as the Str\"omgren $v$ data only.

Like all previous authors, we identify the strongest mode as radial. The 
theoretically predicted amplitude ratios with respect to $u$ are 
systematically higher than observed. We do not believe that this is a 
normalisation error because the measured $u$ and $U$ amplitudes agree 
quite well. We rather think that this is due to an imperfect choice of 
model parameters and because we ignored the chemical peculiarity of the 
atmosphere. Nevertheless, the data strongly suggest that the dominant 
mode is radial.

The photometric amplitudes of the two other modes indicate they are of 
rather high and even spherical degree, most likely $l=4$ or 6. Our 
neglect of convective core overshoot does not affect these mode 
identifications. Whereas overshooting would modify the deeper interior 
structure of the star, the photometric amplitude ratios are mostly 
determined in the stellar photosphere.

Neiner et al.\ (2003) identified the 6.822\,\cd mode as either $l=3$ or 
4. Therefore, the combined evidence points towards an $l=4$ mode, which 
is also the more likely identification for the 7.757\,\cd signal because 
of reasons of geometrical cancellation (Daszy{\'n}ska-Daszkiewicz et 
al.\ 2002). This result is supported by Briquet et al.\ (2012), who 
could spectroscopically constrain the azimuthal order of the modes and 
who also discuss V2052~Oph more deeply in terms of convective core 
overshooting.

\subsection{V986 Oph}

For this star, no spectroscopic temperature and surface gravity 
determinations are available in the literature. The GCPD contains 
Str\"omgren colour indices for V986 Oph, but no Geneva indices. However, 
we have our own photometry in this system available. In Table 4, we 
first compare the standard Geneva colours for V2052 Oph from the GCPD to 
those obtained from our data, demonstrating that they agree within a few 
millimagnitudes. Consequently, we trust the values that we obtained for 
V986 Oph.

\begin{table*}
\begin{minipage}{112mm}
\caption[]{Geneva visual magnitudes and colours$^1$ for our target stars.}
\begin{center}
\begin{tabular}{lccccccc}
\hline
Star & $VM$ & $U$ & $V$ & $B1$ & $B2$ & $V1$ & $G$\\
\hline
V2052 Oph (literature) & 5.803 & 0.598 & 0.855 & 0.834 & 1.543 & 1.563 & 2.018\\
V2052 Oph (this work) & 5.805 & 0.594 & 0.856 & 0.832 & 1.543 & 1.561 & 2.014\\
V986 Oph (this work) & 6.119 & 0.268 & 0.984 & 0.791 & 1.584 & 1.682 & 2.163\\
\hline
\end{tabular}
\end{center}
\smallskip
$^1 VM$ is the visual magnitude, whereas the other parameters are colour 
indices with respect to the $B$ band magnitude (Golay 1972).
\end{minipage}
\end{table*}

Using the standard values in the two photometric systems as input for
photometric calibrations, we must proceed with caution. First, the Geneva
colours of the star are somewhat out of the range of the calibrations by
K\"unzli et al.\ (1997). Extrapolating their grids, one arrives at $T_{\rm
eff} \approx 36000$~K, log~$g\approx 3.8$. Second, the mean Str\"omgren
colours listed in the GCPD and the calibrations implemented by Napiwotzki
et al. (1993) yield $T_{\rm eff} \approx 35600$~K, log~$g \approx 3.0$.  
The latter values however imply a stellar mass in excess of $40
M_{\odot}$, inconsistent with its B0IIIn spectral type. More reliable
seems to be the $T_{\rm eff} \approx 34700$~K value from the calibration
of the $[u-b]$ index by Napiwotzki et al.\ (1993).

Daszy{\'n}ska-Daszkiewicz (2001) determined $T_{\rm eff} = 30100 \pm 
2300$\,K, log~$g=4.0 \pm 0.5$, $[m/H]=0.0$, and $E(B-V)=0.228\pm0.029$ 
from IUE and visual fluxes. The value for reddening is consistent with 
$E(b-y)=0.203$ from Str\"omgren photometry, but the effective 
temperature is much lower, and the surface gravity higher than from the 
photometric calibrations. These parameters rather imply a $16 M_{\odot}$ 
star, but the large error bar on log~$g$ would allow masses up to $25 
M_{\odot}$.

For the purpose of example, we continue with $T_{\rm eff} = 
34700\pm1400$~K, log~$g = 3.8\pm0.3$. We proceeded similar as we did for 
V2052~Oph, computing theoretical photometric amplitudes of the $0 \leq l 
\leq 7$ modes, but for models with masses between 25 and 36 $M_{\sun}$ 
in steps of 1 $M_{\sun}$, and in a temperature range of $4.522 \leq \log 
T_{\rm eff} \leq 4.558$. A frequency range of $2.7 - 3.8$~\cd was 
considered for nonradial modes, and $3.2 - 3.4$~\cd for radial modes (to 
restrict the number of possible radial overtones). The comparison 
between the observed and computed amplitude ratios and a $\chi^2$ 
analysis are shown in Fig.\ 7, where we have restricted ourselves to the 
Str\"omgren data because the oscillation was not present at a 
significant level in the Geneva measurements.

\begin{figure}
\includegraphics[width=85mm,viewport=00 02 260 483]{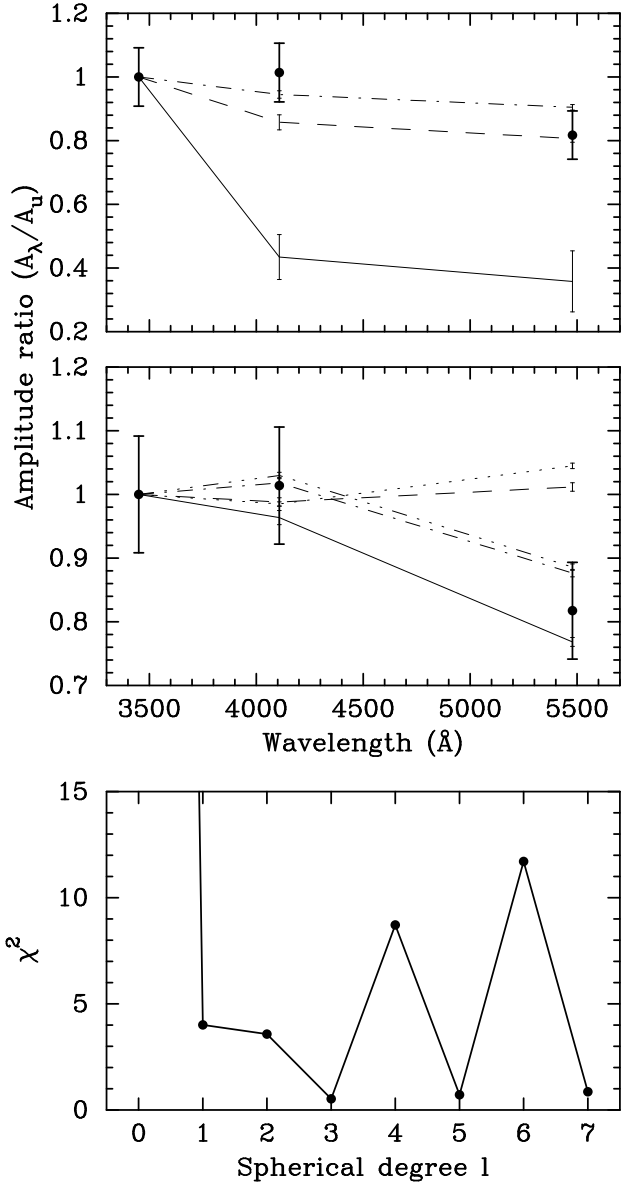}
\caption[]{Mode identifications for V986 Oph from a comparison of observed
and theoretical $uvy$ amplitude ratios, normalised at $u$. In the upper
two panels, the filled circles with error bars are the observed amplitude
ratios. The full line in the uppermost panel is the theoretical
predictions for radial modes, the dashed line for dipole modes, and the
dashed-dotted lines for quadrupole modes. In the middle panel, the full
line is for $l=3$ modes, the dashed line for $l=4$, the dashed-dotted
lines for $l=5$, the dotted lines for $l=6$, and the 
dashed-triple-dotted lines are for $l=7$ modes. The thin error bars
denote the uncertainties in the theoretical amplitude ratios. The lowest
panel shows the results of a $\chi^2$ analysis of the amplitudes.}
\end{figure}

The results of this process clearly argue against a radial pulsation 
mode. Considering nonradial modes, the observed amplitude ratios imply 
that the dominant signal in the light curve is most likely due to an 
$l=3, 5$ or 7 mode. The lowest $\chi^2$, but also the smallest 
geometrical cancellation, then favour an identification as $l=3$, if 
taken at face value. Adopting a lower mass, as implied by the $T_{\rm 
eff}/\log g$ values by Daszy{\'n}ska-Daszkiewicz (2001) results in a 
qualitatively consistent picture with $l=3, 5$ or 7 as the modes best 
reproducing the observed amplitude ratios. We refer to the discussion of 
the credibility of this mode identification near the end of Sect.\ 5.2.

\section{Discussion}

\subsection{V2052 Oph}

As mentioned in the Introduction, the presence of a radial mode in the 
star's pulsation spectrum allows to derive its mean density, provided 
the radial overtone is known. To this end, model evolutionary tracks 
were computed with the Warsaw-New Jersey stellar evolution code, for a 
rotational velocity of 80 \kms on the ZAMS (to match the rotation 
period) and other input parameters as specified in Sect.\ 4.1. 
Nonadiabatic mode frequencies were calculated with the Warsaw pulsation 
code (e.g., see Pamyatnykh et al.\ (1998) for a description of these 
codes), and models sought that had a radial mode at the observed 
frequency. Fig.\ 8 shows the result of this procedure in the form of a 
theoretical HR diagram.

\begin{figure}
\includegraphics[width=85mm,viewport=0 0 303 293]{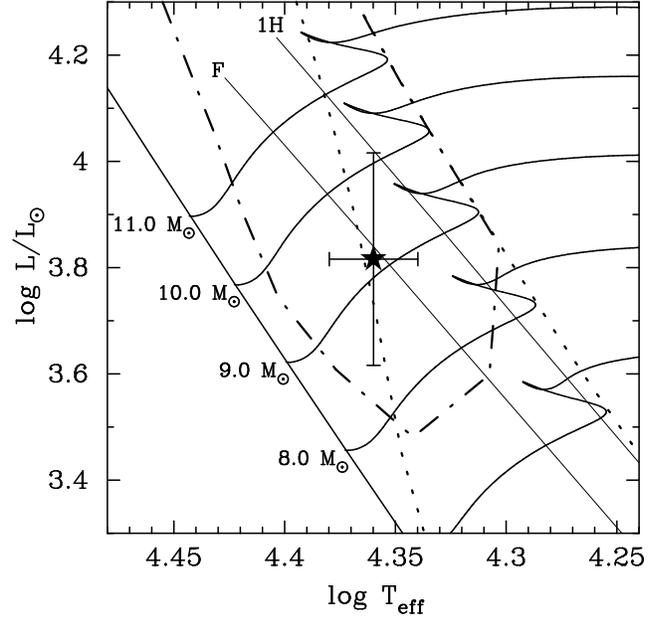}
\caption[]{Constraints on the position of V2052 Oph (star symbol with 
error bars) in the HR Diagram. Some stellar evolutionary tracks are 
plotted and labelled with corresponding masses, and the theoretical 
edges of the $\beta$~Cep (dashed-dotted line) and Slowly Pulsating B 
(SPB) star (dotted line) instability strips (Pamyatnykh \& Ziomek 2007) 
are shown. The thin full lines connect models with radial modes at the 
observed frequency; "F" stands for the fundamental mode, "1H" for the 
first overtone.}
\end{figure}

The effective temperature and luminosity derived for V2052 Oph in Sect.\ 
4.1 is in best agreement with the hypothesis that the radial mode is the 
fundamental, but it cannot be excluded that it is the first overtone. 
In the first case, the stellar radius would be $5.3\pm0.1$\,R$_{\odot}$ 
and the inclination of the rotation axis thus $54\degr <i< 58\degr$, 
given the star's rotation period and $v \sin i$. These values change to 
$6.45\pm0.15$\,R$_{\odot}$ and $42\degr <i< 44\degr$ if the radial mode 
was the first overtone. Unfortunately, the scarcity of additional 
pulsation modes and their unknown azimuthal order leave meager prospects 
for asteroseismic constraints other than deriving the mean stellar 
density.

Attempting to fit the $l=4$ mode frequencies with models of the same 
radial mode period but varying mass gave a number of possible solutions. 
Unsurprisingly, if the radial mode was the fundamental, models of lower 
mass appear more likely because these are more evolved (cf.\ Fig.\ 8) 
and therefore have more mixed modes of $l=4$. If the radial mode was 
assumed to be the first overtone, no such preference was seen.

The only statement we can make is that the two $l=4$ modes are unlikely 
to be rotationally split m-modes of the same radial overtone unless 
allowing for differential interior rotation. The same conclusion was 
reached by Briquet et al.\ (2012), with their independent model approach 
and identifications of $m$. The reason for this finding is that the 
effect of the Coriolis and centrifugal forces on the rotational 
frequency splitting are very similar at the given rotation rate and for 
$l=4$, no matter whether the mode under consideration is a p, g or mixed 
mode.

Neiner et al.\ (2003) determined a very precise rotation period for the 
star despite a non-optimal temporal distribution of their data in terms 
of annual aliasing problems. The rotation period we have obtained is 
consistent within the errors with the more precise one by Neiner et al.\ 
(2003), and accurate enough to rule out that their value is affected by 
aliases. Therefore we confirm 3.638833~d as the best available rotation 
period of V2052~Oph.

In Fig.\ 4, the phase of minimum rotational light variation coincides 
with minimum magnetic field strength and minimum UV spectral line 
equivalent width, as determined by Neiner et al.\ (2012a). In $B1, v$ 
and $B$ there is also a double maximum, as in the UV line strength. The 
shape of these variations indicates that both magnetic poles are seen 
during a rotation cycle, i.e. the sum of the angles of the inclination 
of the rotation axis and the magnetic obliquity $i+\beta$ must exceed 
$90\degr$.

Neiner et al.\ (2012a) determined $i$ to be $53\degr<i<77\degr$ from 
modelling Stokes profiles. This is consistent with the inclination of 
the rotation axis we obtained with a stellar radius corresponding to 
radial fundamental mode pulsation, but not with the value assuming the 
radial mode is the first overtone. We therefore rule out the latter 
possibility. Using $r=B_{\rm min}/B_{\rm 
max}=\cos(i-\beta)/\cos(i+\beta)$, where $r$ is the ratio of the minimum 
and maximum magnetic field strength of the rotation cycle, and the $r$ 
values determined by Neiner et al.\ (2012a), we obtain 
$58\degr<\beta<66\degr$.

As mentioned, the light curve of V2052 Oph shows rotational modulation 
(Fig.\ 4) in phase with the magnetic field and UV wind variations. This 
could be due to spots at the surface of the star (such as those 
suggested by Neiner et al.\ 2012a) or magnetically confined clouds in 
the circumstellar environment (e.g.\ Townsend \& Owocki 2005). Using the 
magnetic field value ($B_{\rm pol} = 400$~G, Neiner et al.\ 2012a), the 
wind velocity estimated in UV data ($v_{\rm inf} = 500$\,\kms, Neiner et 
al.\ 2003), the stellar parameters from Sect.\ 4.1 and above ($M = 9.2$ 
M$_\odot$, $R = 5.3$ R$_\odot$, $v \sin i = 61$\,\kms, $i = 56^\circ$), 
and a mass loss typical of a B2 star ($\dot{M} = 10^{-9}$ 
M$_\odot$~yr$^{-1}$), we derived the magnetic confinement parameter 
$\eta_*$ (see ud-Doula \& Owocki 2002) of V2052 Oph.

We obtained that $\eta_* = 2576$, the Alfven radius is $R_A = 7.12 R$, 
and the Kepler radius is $R_K = 3.05 R$. This implies that magnetic 
confinement should occur ($\eta_* > 1$) at the magnetic equator between 
$R_K$ and $R_A$. Indeed in this region wind particles get trapped in 
closed field loops and remains centrifugally supported. Above $R_A$ 
material escapes as the wind streches field lines open. Below $R_K$ 
material lacks sufficient centrifugal support and falls back onto the 
star, but this transient material can still create a dynamical 
magnetosphere (ud-Doula, Owocki \& Townsend 2008, Petit et al.\ 2011).

However, a centrifugally supported magnetosphere usually produces 
H$\alpha$ emission and such emission has never been observed in V2052 
Oph. Moreover, Oskinova et al.\ (2011) showed that V2052 Oph is only 
very weakly X-ray luminous. A centrifugally supported magnetosphere 
could exist without producing H$\alpha$ or much X-ray emission if the 
density or temperature of the wind was not appropriate or if the 
confinement timescale was too long (see, e.g., Neiner et al.\ 2012b for 
a more detailed discussion of the emission measure in magnetospheres).

V2052 Oph is not the only magnetic $\beta$~Cep star known. Telting, 
Aerts \& Mathias (1997) spectroscopically detected rotationally equally 
split frequencies around the dominant radial pulsation mode of 
$\beta$~Cep itself, and discussed whether these could be temperature 
spots on the surface or due to a magnetically distorted oblique 
pulsation mode (the magnetic field was reported by Henrichs et al.\ 
2000). On the other hand, no such rotational frequency splitting has 
been reported for $\xi^1$~CMa (Saesen, Briquet \& Aerts 2006, 
Fourtune-Ravard et al.\ 2011). Both stars have a single or dominant 
radial mode, such as V2052 Oph. For our target, we find no signals at 
frequencies split by one or two times the rotation frequency around the 
radial pulsation, within a limit of 0.2 mmag in amplitude.

\subsection{V986 Oph}

In Sect.\ 3.2 we remarked that the residual scatter in our light curves 
prewhitened by the single coherent frequency is considerably higher than 
that in the other time series from this campaign. Since the two 
comparison stars are farther apart from each other on the sky as V986 
Oph is from either of them, this cannot be due to residual data 
reduction errors. Furthermore, the scatter in colour light curves of 
V986 Oph, e.g. $u-y$, is much less than the residual scatter in the 
individual passbands, and virtually the same as in the differential 
comparison star data in the same filter combination. We therefore 
conclude that this high apparent scatter in the light curves actually 
represents intrinsic variability of V986 Oph, and that this kind of 
variability is not dominated by changes in the stellar effective 
temperature.

The main variability frequency we found is consistent with the one in 
the 1987 data by Cuypers et al.\ (1989), and the amplitude is 
comparable. However, these authors also remarked on the presence of 
longer-term light variations on time scales longer than half a day. We 
do not find coherent variability on such time scales in the complete 
data set. We therefore subdivided our measurements into chunks 
comparable to the extent of the data by Cuypers et al.\ (1989) and 
analysed the residuals after prewhitening the main periodicity. 
Longer-term variability is present, but we find nothing periodic. 
Unfortunately, our data have too low a duty cycle for invoking 
techniques such as Time-Fourier analysis to search for some possible 
short-lived periodic variations.

V986 Oph is not the first case of a $\beta$~Cep star for which 
unknown additional variability besides that of pulsational origin has 
been found. Jerzykiewicz (1978) and Handler et al.\ (2006) discussed 
this problem for the star 12 (DD) Lac (11.5 M$_{\odot}$), Jerzykiewicz 
et al.\ (2005) for $\nu$~Eridani (9.6 M$_{\odot}$), and Handler et al.\ 
(2005) for $\theta$\,Oph ($\sim 8.5$ M$_{\odot}$), where it seems 
present at a lower level. On the other hand, for V2052 Oph we did not 
find such evidence, and it is hardly, if at all present in the MOST 
photometry of $\gamma$~Peg (8.5 M$_{\odot}$, Handler et al.\ 2009). One 
might therefore speculate that the more massive the star, the stronger 
this additional variability.

In this context it is very interesting that Blomme et al.\ (2011) 
analysed CoRoT light curves of three O stars and also found some 
apparently incoherent variability, in all three targets. These authors 
suggested that it could be due to subsurface convection, granulation or 
wind variability. On the other hand, Balona et al.\ (2011) suggested 
that the low frequencies observed in the amplitude spectra of Kepler 
B-type stars are due to many simultaneous gravity mode oscillations with 
high spherical degree. The present data for V986 Oph do not allow us to 
distinguish between those possibilites, and also not to argue against 
pulsation in modes of high spherical degree, because those would not 
generate strong colour variability (cf.\ Daszy{\'n}ska-Daszkiewicz \& 
Pamyatnykh 2012). However, we do point out that this presently 
unexplained variability may occur in stars with masses down to 
9\,M$_{\odot}$.

Some authors have (e.g., Jerzykiewicz 1975) questioned the 
membership of V986 Oph to the class of $\beta$~Cep stars due to its long 
variability period. Because the star rotates rapidly, several 
possibilities need to be considered. With the temperature and luminosity 
estimate for V986 Oph from the photometric data in Sect.\ 4.2 the star 
would have a radius of $12\pm4$\,R$_{\odot}$, which yields a rotation 
period of about two days assuming $v_{rot} = 300$\,\kms. As the critical 
(break-up) rotational velocity of such massive stars is around 
400\,\kms, the rotation frequency cannot exceed 0.7\,\cd. For the 
possibility of a $\sim 16$\,M$_{\odot}$, star, this upper limit 
increases to 2.7\,\cd. Therefore we rule out that the single coherent 
signal we found in the light curves of V986 Oph is due to rotation.

Another hypothesis would be a g mode frequency rotationally split into 
the p/mixed mode domain. However, because of the large uncertainties in 
the stellar mass and effective temperature, we cannot reach a conclusion 
for this possibility and stay with the assumption that V986 Oph is a 
$\beta$~Cephei star.

The frequency of the single coherent variability signal has changed from 
the first published observations, as summarized by Jerzykiewicz (1975). 
Up to his paper, the frequency was quoted as 3.44\cd or somewhat 
higher. Later, Fullerton et al.\ (1985) gave two different periods for 
the different seasons 1980 and 1984, the latter consistent with the 
3.29\cd frequency determined by Cuypers et al.\ (1989) and us. Therefore 
this frequency must have changed some time in the 1980's, by an amount 
too large to be explicable by stellar evolution. Most likely, it is just 
due to a change of the dominant pulsation mode of the star, which has 
been observed in at least one other $\beta$~Cep star before 
(Jerzykiewicz \& Pigulski 1996).

Concerning the amplitude, the published light range is of the order of 
0.02 to 0.03 mag. This is comparable to what we see in our data (cf.\ 
Fig.\ 1). However, the amplitude of the main periodicity may have 
dropped, or the larger values reported in the literature, based on much 
smaller data sets, are biased by the incoherent variability. 
Jerzykiewicz (1975) already remarked on the unusually low $U/B$ 
amplitude ratio from the viewpoint of $\beta$~Cep pulsation, as 
manifested in the data of Hill (1967). Our observed $u/v$ amplitude 
ratio is consistent with that, and can be best explained with an odd-l 
mode of fairly high degree ($l\geq3$).

Such a mode identification is unexpected for two reasons. First, 
geometrical cancellation is very strong for modes with odd spherical 
degree larger than one (Daszy{\'n}ska-Daszkiewicz et al.\ 2002). Second, 
for such a rapidly rotating star one would naively expect a preference 
for $l=2$ modes due to the distortion of the stellar shape. However, as 
demonstrated by Townsend (2003), our implicit assumption that the 
geometry of the pulsation mode of V986 Oph can be described in the form 
of a single spherical harmonic may not be correct. Also, the photometric 
amplitudes of rapidly rotating stars depend on the azimuthal order of 
the modes as well as the aspect under which they are viewed (Townsend 
2003).

Our most likely mode identification as $l=3$ is therefore
uncertain and calls for a high-resolution spectroscopic investigation. 
This would also serve to derive more reliable values of the stellar 
temperature and surface gravity than we have available. In particular, 
it would be interesting to confirm or reject the high stellar mass we 
inferred.

\section{Conclusions}

In an attempt to understand the pulsational behaviour of $\beta$~Cep 
stars that are more complicated than those previously studied with 
asteroseismic methods (slow rotators with fairly large amplitudes), we 
have carried out an extensive multisite campaign. Results from the 
photometric investigation were however insufficient to perform a 
detailed asteroseismic study, as only three pulsation modes were 
detected for V2052 Oph, and one for V986 Oph.

However, it is interesting that the nonradial modes present are of 
higher spherical degree than commonly found. We are aware of only a few 
cases with observationally identified $l=3$ (e.g., Briquet et al.\ 2009) 
or $l=4$ modes (e.g., Aerts, Waelkens \& De Pauw 1994), and these stars 
tend to rotate more rapidly than those dominated by modes of low 
spherical degree. 
 
Therefore the often-made assumption that $\beta$~Cep pulsation modes 
detected photometrically from the ground are $l\leq2$ needs to be 
questioned. In the context of highly sensitive space photometry this 
assumption is of course even more doubtful. Since the amplitude 
reduction due to geometrical cancellation of sufficiently high-l 
pulsation modes only goes as $\sim l^{-1/2}$ (e.g., 
Daszy{\'n}ska-Daszkiewicz et al.\ 2002), the gain in the number of 
oscillation frequencies detected is offset by the larger uncertainty in 
mode typing.

Photometric identifications of modes with high spherical degree become 
largely degenerate for even and odd modes with $l \geq 3$. This 
situation can be relieved by obtaining simultaneous spectroscopy. The 
amplitude ratios and phase shifts between the radial velocity and light 
curves allow some separation between high-l modes 
(Daszy{\'n}ska-Daszkiewicz \& Pamyatnykh 2012), and of course 
line-profile variations offer a multitude of possibilities for 
identifying modes (e.g., Telting 2008 and references therein).

It therefore seems that if we are to understand the interior structure 
of more rapidly rotating $\beta$~Cep stars than those studied so far, 
all observationally and theoretically available tools need to be 
exploited. Photometric measurements need to be made in multiple 
passbands, and (simultaneous) spectroscopic observations must be 
acquired. The interpretation of these data may require the inclusion of 
the effects of rotation on a star-to-star basis because observables can 
be affected to the extent that mode identifications, a prerequisite for 
asteroseismology, may be erroneous (see Townsend 2003). However, to shed 
light on important astrophysical problems, such as internal angular 
momentum transport, such concerted efforts will be worthwhile.

\section*{ACKNOWLEDGEMENTS}

This work has been supported by the Austrian Fonds zur F\"orderung der 
wissenschaftlichen Forschung under grant R12-N02. KU acknowledges 
financial support by the Spanish National Plan of R\&D for 2010, project 
AYA2010-17803. GH thanks Jadwiga Daszy{\'n}ska-Daszkiewicz and Patrick 
Lenz for helpful discussions, and Luis Balona for permission to use his 
software. EG acknowledges support from the Austrian Science Fund (FWF), 
project number P19962-N16. MB is a F.R.S.-FNRS Postdoctoral Researcher, 
Belgium.

This work is based in part on observations made with the Mercator 
Telescope, operated on the island of La Palma by the Flemish Community, 
at the Spanish Observatorio del Roque de los Muchachos of the Instituto 
de Astrof\'{\i}sica de Canarias, and at the South African Astronomical 
Observatory.

\bsp


\begin{thebibliography}{99}

\bibitem[]{}Abt H.~A., Levato H., Grosso M. 2002, ApJ 573, 359

\bibitem[]{}Aerts C., Waelkens C., de Pauw M., 1994, A\&A 286, 136

\bibitem[]{}Aerts C., Christensen-Dalsgaard J., Kurtz D. W. 2010,
Asteroseismology, (Springer-Verlag, Berlin)

\bibitem[]{}Aerts C., et al. 2003, Science, 300, 1926

\bibitem[]{}Aerts C., Briquet M., Degroote P., Thoul A., van Hoolst T., 
2011, A\&A 534, 98

\bibitem[]{}Asplund M., Grevesse N., Sauval A. J., Allende Prieto C.,
Kiselman D., 2004, A\&A 417, 751

\bibitem[]{}Balona L. A., Shobbrook R. R., 1974, MNRAS 211, 375

\bibitem[]{}Balona L. A., Evers E. A., 1999, MNRAS 302, 349

\bibitem[]{}Balona L. A., et al., 2011, MNRAS 413, 2403

\bibitem[]{}Blomme R., et al., 2011, A\&A 533, A4

\bibitem[]{}Breger M., et al., 1993, A\&A 271, 482

\bibitem[]{}Breger M., et al., 1999, A\&A 349, 225

\bibitem[]{}Briquet M., et al., 2009, A\&A 506, 269

\bibitem[]{}Briquet M., et al., 2012, MNRAS, to be submitted

\bibitem[]{}Crawford D. L., 1978, AJ 83, 48 

\bibitem[]{}Cugier H., Dziembowski W. A., Pamyatnykh A. A., 1994, A\&A
291, 143

\bibitem[]{}Cuypers J., Balona L.~A., Marang F. 1989, A\&AS, 81, 151

\bibitem[]{}Daszy{\'n}ska-Daszkiewicz J., 2001, PhD thesis, University 
of Wroc{\l}aw

\bibitem[]{}Daszy{\'n}ska-Daszkiewicz J., Pamyatnykh A. A., 2012, in 
{\it Impact of new instrumentation and new insights in stellar 
pulsations} eds.\ L.\ A.\ Balona et al., Astrophys.\ Space Sci.\ Proc.\ 
series, in press (arXiv:1112.2572)

\bibitem[]{}Daszy{\'n}ska-Daszkiewicz J., Dziembowski W. A.,
Pamyatnykh A. A., Goupil M.-J., 2002, A\&A 392, 151

\bibitem[]{}Desmet M., et al., 2009, MNRAS 396, 1460


\bibitem[]{}Flower P.\ J., 1996, ApJ 469, 355

\bibitem[]{}Fourtune-Ravard C., Wade G.\ A., Marcolino W.\ L.\ F., 
Shultz M., Grunhut J.\ H., Henrichs H.\ F., in {\it Active OB stars: 
structure, evolution, mass loss, and critical limits}, Proc.\ IAU Symp.\ 
272, p.\ 180

\bibitem[]{}Frost E. B., 1902, ApJ, 15, 340

\bibitem[]{}Fullerton A. W., Bolton C. D., Penrod G. D., 1985, JRASC 79, 
236

\bibitem[]{}Golay M., 1972, Vistas in Astronomy 14, 13

\bibitem[]{}Handler G., et al., 2006, MNRAS 365, 327

\bibitem[]{}Handler G., et al., 2009, ApJ 698, L56

\bibitem[]{}Handler G., in {\it Planets, Stars and Stellar Systems}, 
eds.\ T. D. Oswalt et al., Springer-Verlag, Berlin, 2012, in press 
(Chapter 27)

\bibitem[]{}Handler G., Shobbrook R. R., Mokgwetsi T., 2005, MNRAS 362,
612

\bibitem[]{}Hauck B., Mermilliod M., 1998, A\&A 129, 431

\bibitem[]{}Henrichs H. F., et al., 2000, in {\it The Be Phenomenon in 
Early-Type Stars}, eds.\ M. A. Smith \& H. F. Henrichs, ASP Conf.\ Ser.\ 
214, p.\ 324

\bibitem[]{}Heynderickx D., Waelkens C., Smeyers P., 1994, A\&AS 105, 447

\bibitem[]{}Hill G., 1967, ApJS 14, 263

\bibitem[]{}Jerzykiewicz M., 1972, PASP 84, 718

\bibitem[]{}Jerzykiewicz M., 1975, Acta Astr.\ 25, 81

\bibitem[]{}Jerzykiewicz M., 1978, Acta Astr. 28, 465

\bibitem[]{}Jerzykiewicz M., 1993, A\&AS 97, 421

\bibitem[]{}Jerzykiewicz, M., Pigulski, A., 1996, MNRAS, 282, 853

\bibitem[]{}Jerzykiewicz M. et al. 2005, MNRAS 360, 619

\bibitem[]{}Jones D. H. P., Shobbrook R. R., 1974, MNRAS 166, 649

\bibitem[]{}K\"unzli M., North P., Kurucz R. L., Nicolet B., 1997, A\&AS
122, 51

\bibitem[]{}Ledoux P., 1951, ApJ 114, 373

\bibitem[]{}van Leeuwen F., 2007, A\&A 474, 653

\bibitem[]{}Lenz P., Breger M., 2005, Comm. Asteroseism. 146, 53

\bibitem[]{}Mermilliod J.-C., Mermilliod M., Hauck B., 1997, A\&AS 124, 
349

\bibitem[]{}Montgomery M.\ H., O'Donoghue D., 1999, Delta Scuti Star
Newsletter 13, 28 (University of Vienna)

\bibitem[]{}Morel T., Butler K., Aerts C., Neiner C., Briquet M., 2006, 
A\&A 457, 651

\bibitem[]{}Napiwotzki R., Sch\"onberner D., Wenske V., 1993, A\&A 268,
653

\bibitem[]{}Neiner C., et al., 2003, A\&A 411, 565

\bibitem[]{}Neiner C., et al., 2012a, A\&A 537, A148

\bibitem[]{}Neiner C., et al., 2012b, A\&A, in press

\bibitem[]{}Niemczura E., Daszy{\'n}ska-Daszkiewicz J., 2005, A\&A 433,
659

\bibitem[]{}Pamyatnykh A. A., Ziomek, W., 2007, Comm. Asteroseism. 150, 
207

\bibitem[]{}Pamyatnykh A. A., Dziembowski W. A., Handler G., Pikall H.,
1998, A\&A 333, 141

\bibitem[]{}Pamyatnykh A. A., Handler G., Dziembowski W. A., 2004, MNRAS 
350, 1022

\bibitem[]{}Petit V., et al., 2012, in {\it Four Decades of Massive Star 
Research}, ed.\ L.\ Drissen, ASP Conf.\ Ser., in press (arXiv:1111.1238)

\bibitem[]{}Pigulski A., Pojma{\'n}ski G., 2008, A\&A 477, 917

\bibitem[]{}Saesen S., Briquet M., Aerts C., 2006, Comm. Asteroseism. 
147, 109

\bibitem[]{}Seaton M. J., 2005, MNRAS 362, L1

\bibitem[]{}Stankov A., Handler G., 2005, ApJS 158, 193

\bibitem[]{}Stateva I., Niemczura E., Iliev I., 2010, Pub.\ Astr.\ Obs.\ 
Belgrade 90, 179


\bibitem[]{}Telting J. H., 2008, Comm. Asteroseism., 157, 112

\bibitem[]{}Telting J. H., Aerts C., Mathias, P., 1997, A\&A 322, 493

\bibitem[]{}Telting J. H., Schrijvers C., Ilyin I. V., Uytterhoeven K., De
Ridder J., Aerts C., Henrichs H. F., 2006, A\&A 452, 945

\bibitem[]{}Townsend R.\ H.\ D., 2003, MNRAS 343, 125

\bibitem[]{}Townsend R.\ H.\ D., Owocki S. P., 2005, MNRAS 357, 251

\bibitem[]{}ud-Doula A., Owocki S. P., 2002, ApJ 576, 413

\bibitem[]{}ud-Doula A., Owocki S. P., Townsend R. H. D., 2008, MNRAS, 
385, 97

\end{thebibliography}
\end{document}